# Digital resolution enhancement in surface plasmon microscopy

I.I. Smolyaninov [1)]*, J. Elliott [2)], G. Wurtz [2)], A.V. Zayats [2)], C.C. Davis [1)]

[1)] Department of Electrical and Computer Engineering, University of Maryland, College Park, MD 20742, USA

[2)] School of Mathematics and Physics, The Queen's University of Belfast, Belfast BT7 1NN, UK

* Phone: 301-405-3255, fax: 301-314-9281, e-mail: smoly@eng.umd.edu

**Abstract**

The use of photonic crystal and negative refractive index materials is known to improve resolution of optical microscopy and lithography devices down to 80 nm level. Here we demonstrate that utilization of well-known digital image recovery techniques allows us to further improve resolution of optical microscope down to 30 nm level. Our microscope is based on a flat dielectric mirror deposited onto an array of nanoholes in thin gold film. This two-dimensional photonic crystal mirror may have either positive or negative effective refractive index as perceived by surface plasmon polartions in the visible frequency range. The optical images formed by the mirror are enhanced using simple digital filters.

**PACS:** 73.20.Mf, 42.70.Qs, 07.60.Pb

Optical microscopy is invaluable for studies of materials and biological entities. However, the spatial resolution of the conventional microscopy is limited by the diffraction of light waves to a value of the order of 200 nm. An order of magnitude better resolution is necessary in order to visualize viruses, proteins, DNA molecules and many other samples. Very recently Pendry [1] suggested that negative refractive index materials may be used to achieve such resolution without resorting to scanning near-field microscopy techniques [2], which are sequential and very slow in its nature. Indeed, ~ 80 nm resolution has been observed in recent optical lithography [3] and microscopy [4] experiments utilizing negative refractive index materials based on surface plasmon polariton (SPP) [5] optics (it should be noted that resolution enhancement in plasmon polariton microscopy experiments [4] was not limited only to negative refractive index case: resolution appears to be enhanced also in a more general two-dimensional (2D) photonic crystal case with either positive or negative effective refractive index). Even though impressive, ~80 nm resolution is not sufficient for most biological applications. In this paper we demonstrate that utilization of well-known digital image recovery techniques allows us to further improve resolution of the SPP-based optical microscope down to 30 nm level.

Our microscope is based on a flat dielectric mirror deposited onto an array of nanoholes in a thin gold film (Fig.1). A detailed account of the microscope design and its performance can be found in ref. [4]. The photonic crystal mirror used in the microscope may have either negative or positive effective refractive index as perceived by surface plasmon polartions in the visible frequency range. The case shown in Fig.1 corresponds to the case of negative refractive index. The optical images of various test patterns consisting of differently-shaped nanoholes in gold films were obtained at illuminating laser wavelengths of 502 and 515 nm, and magnified by the photonic crystal mirror and formed over the flat gold film surface (as shown in Fig.1). The resulting optical images were observed due to SPP scattering using regular optical microscope. The sign of the effective refractive index of the mirror is defined by the sign of the SPP group velocity near the surface plasmon resonance. In the experiment this sign can be determined by the character of magnification distribution in the SPP-formed images, as described in ref. [4]. Some examples of so produced optical images are shown in Figs.2 and 3.

The spatial resolution of the optical images (the point spread function (PSF) of our microscope) can be measured directly by calculating the cross-correlation between the optical image and the scanning electron microscope (SEM) image of the same nanohole. The results of these calculations in the cases of triplet and U-shaped nanoholes from Figs. 2 and 3 are presented in Fig.4. These calculations demonstrate resolution of the order of PSF~70 nm or ~$\lambda/8$ achieved in these particular imaging experiments. It should be noted that similar resolution of the order of 90 nm at 413 nm laser wavelength was observed recently in focusing experiments with planar parabolic solid immersion mirrors even without the use of surface plasmon polaritons (regular guided modes were used) [6]. Slightly better resolution in our microscopy experiment is due to the fact that SPP wavelength is shorter than the wavelength of guided modes at the same laser frequency. Photonic crystal effects and the effects of negative refraction also play some role in achieving better resolution, as described in ref. [4].

Even though quite an improvement compared to a regular optical microscope, the ~ 70 nm resolution is not sufficient to achieve clear visibility of many nanoholes in the test pattern in Fig.3, which was made in order to emulate variety of shapes of different biological samples. Even though recognizable, most nanoholes appear quite fuzzy. However, the blurring of optical images at the limits of optical device resolution is a very old problem (one may recall the well-publicized recent problem of Hubble telescope repair). One solution of this problem is also well-known. There exists a wide variety of image recovery techniques which successfully fight image blur based on the known PSF of the optical system. One of such techniques is matrix deconvolution based on the Laplacian filter (see Fig.5 for an example). Utilization of such techniques is known to improve resolution by at least a factor of two. However, precise knowledge of the PSF of the microscope in a given location in the image is absolutely essential for this technique to work, since it involves matrix convolution of the experimental image with a rapidly oscillating Laplacian filter matrix (an example of such 5x5 matrix is shown in Fig.5). In our test experiments the PSF of the microscope was measured directly in some particular location of the optical image, as is shown in Fig.4. This measured PSF was used to digitally enhance images of the neighboring nanohole arrays. Similar technique may be used to enhance resolution in the SPP-induced optical images of biological samples, which are measured using the nanohole array background, as described in ref.[4].

Not surprisingly, the use of such digital filters led to approximately two-fold improvement of resolution in the optical images formed by the photonic crystal mirror in both positive and negative effective refractive index cases. This two-fold improvement is demonstrated in Fig.6 for both the triplet and the U-shaped nanoholes shown in Figs 2-4. The point spread function measured as the cross-correlation between the digitally processed optical image and the corresponding SEM image appears to fall firmly into the 30 nm range, which represents improvement of resolution of the SPP-assisted optical microscope down to ~ $\lambda/20$ range. This result may bring about direct optical visualization of many important biological systems.

In conclusion, we have demonstrated quantitatively that the use of photonic crystal mirror may improve resolution of immersion optical microscopy down to 80 nm level. In addition, utilization of well-known digital image recovery techniques allows us to further improve resolution of optical microscope down to 30 nm level. This result may be useful in various biological applications of the SPP-assisted microscope based on photonic crystal mirror.

This work was supported in part by NSF grants ECS-0304046 and CCF-0508213 and by the EPSRC.

**Figure Captions**

**Figure 1.** Schematic view of the surface plasmon microscope (SPM) based on 2D photonic crystal mirror in the case of negative effective refractive index of the mirror as perceived by surface plasmon polaritons.

**Figure 2.** Comparison of the SPP-produced optical (a) and the scanning electron microscope (SEM) images of the test array of triplet nanoholes. Comparison of the Fourier transforms of these images indicates spatial resolution in the optical image of at least 98 nm. This conclusion may be reached from the apparent visibility of higher harmonics of the triplet structure (indicated by the arrows) in the optical image.

**Figure 3.** Scanning electron microscope (SEM) image of an aperiodic nanohole array test sample is shown is (b). The optical image of the test sample obtained using SPPs in the geometry from Fig.1 is shown in (a). The insets in (a) compare the images of a U-shaped nanohole obtained using SPPs and SEM.

**Figure 4.** Calculated cross correlation functions between the SEM and optical images of a triplet nanohole from Fig.2 (top row) and U-shaped nanohole from Fig.3 (bottom row). These calculations indicate the point spread function of the optical microscope of the order of 70 nm.

**Figure 5.** Theoretical modeling of the image recovery using Laplacian filter matrix deconvolution: Laplacian filter (shown in the inset) allows to recover image deterioration due to Gaussian blur, which is evidenced via calculation of the cross correlation of the original SEM image and the image recovered using Laplacian matrix deconvolution method.

**Figure 6.** Calculated cross correlation functions between the SEM and the digitally-enhanced optical images of a triplet nanohole from Fig.2 (top row) and U-shaped nanohole from Fig.3 (bottom row). Comparison of these images with Fig.4 indicates ~ two-fold improvement of the image resolution. Calculated PSF of the digitally-enhanced optical images appears to be of the order of 30 nm.

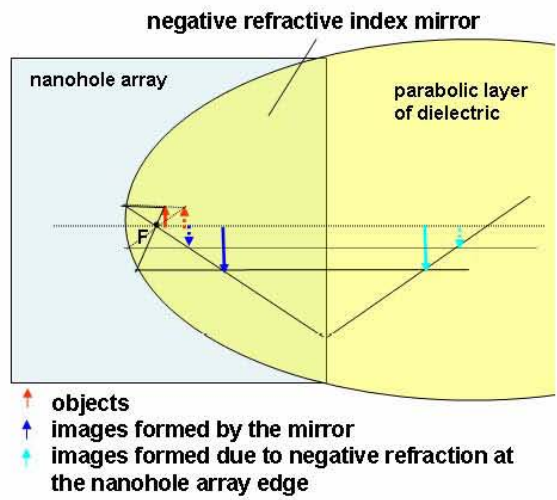

**Fig.1**

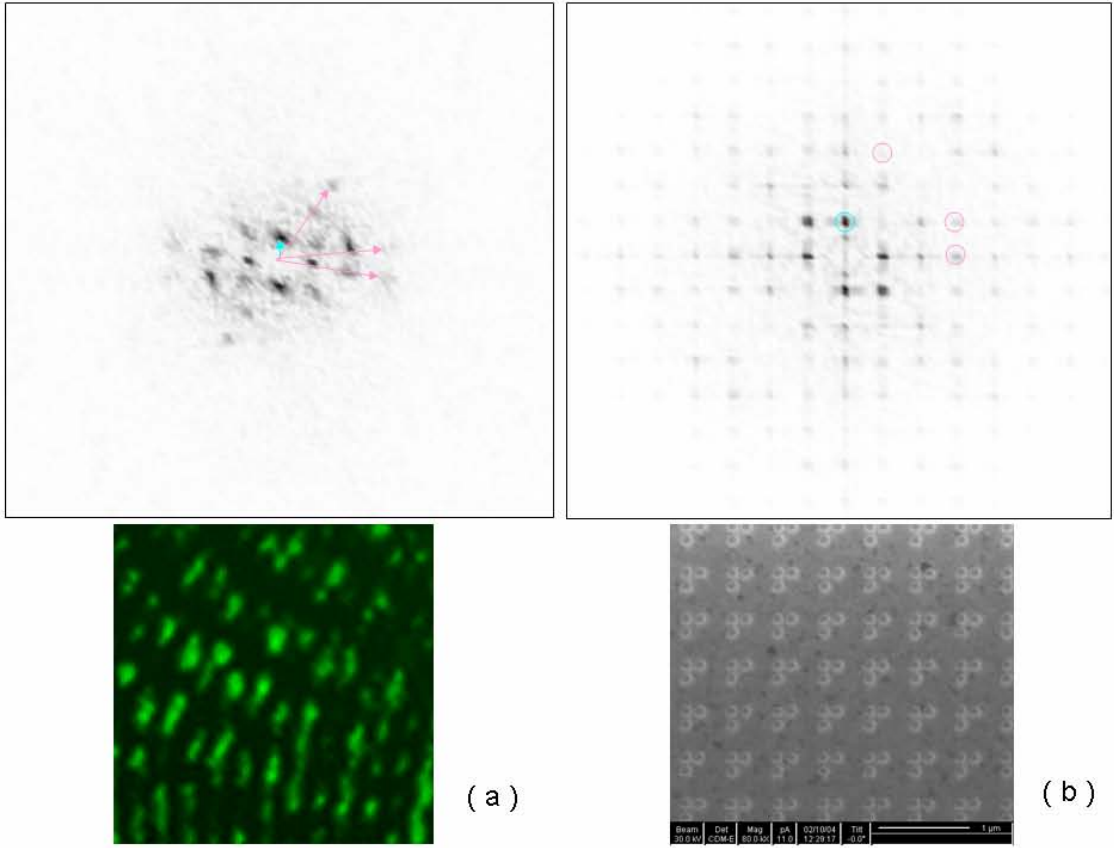

**Fig.2**

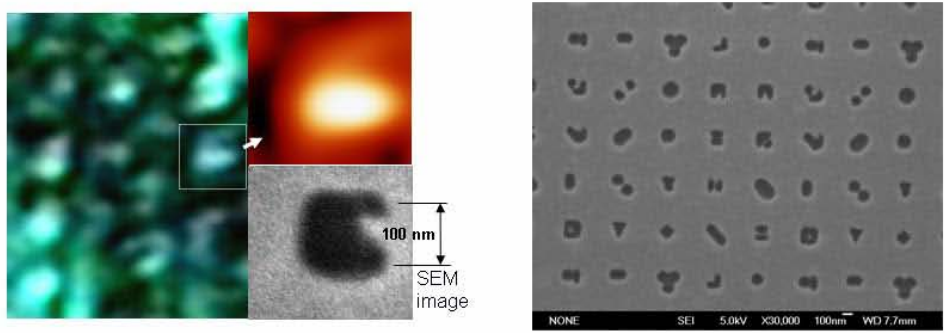

(a)                          (b)

**Fig.3**

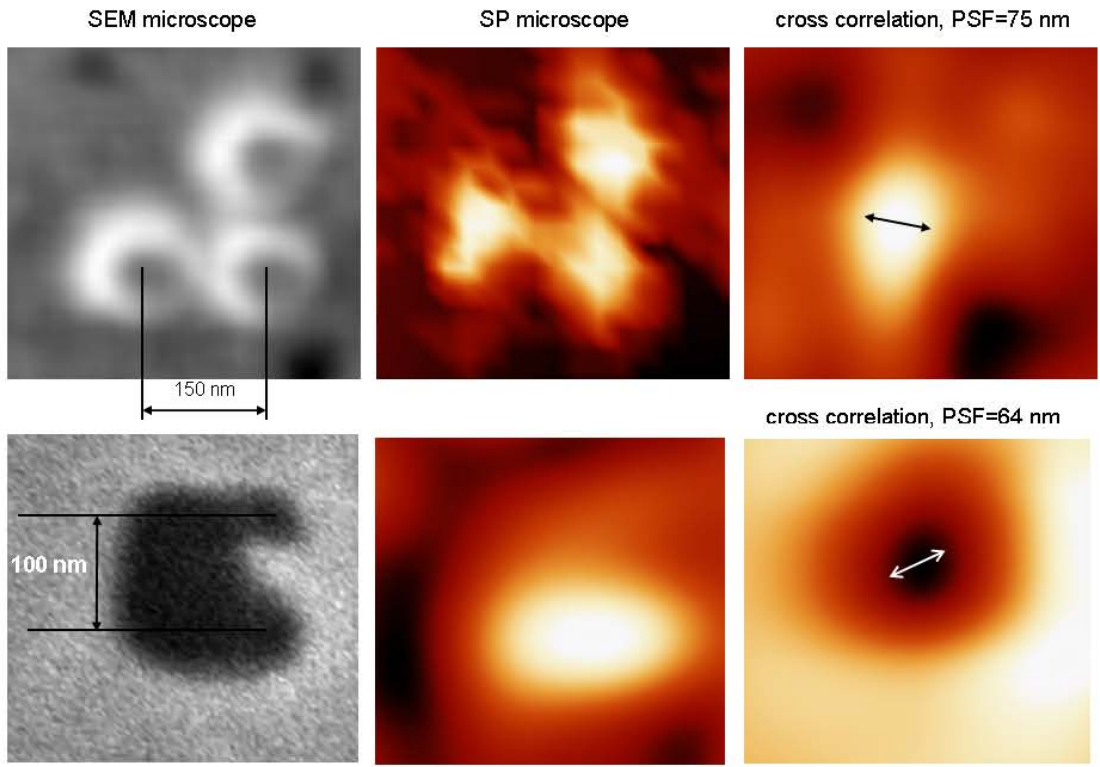

**Fig.4**

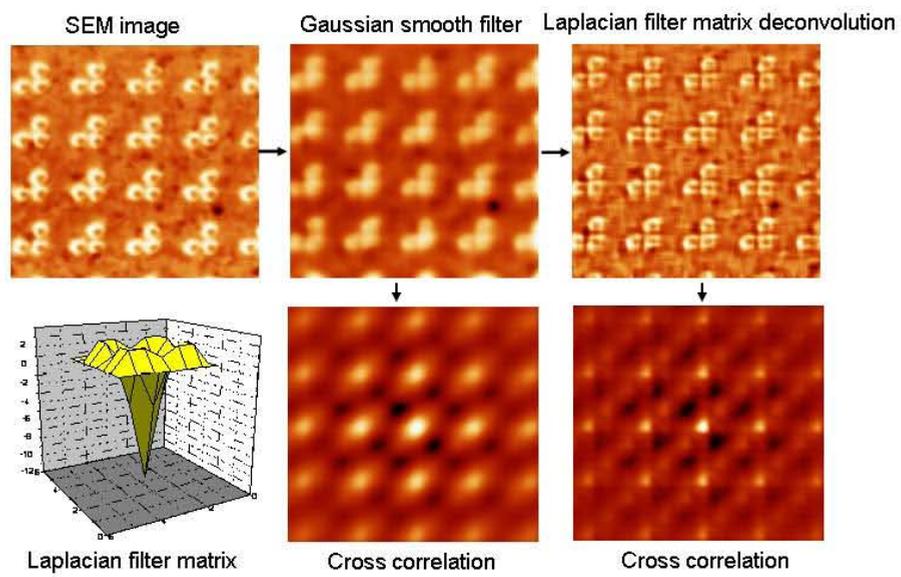

**Fig.5**

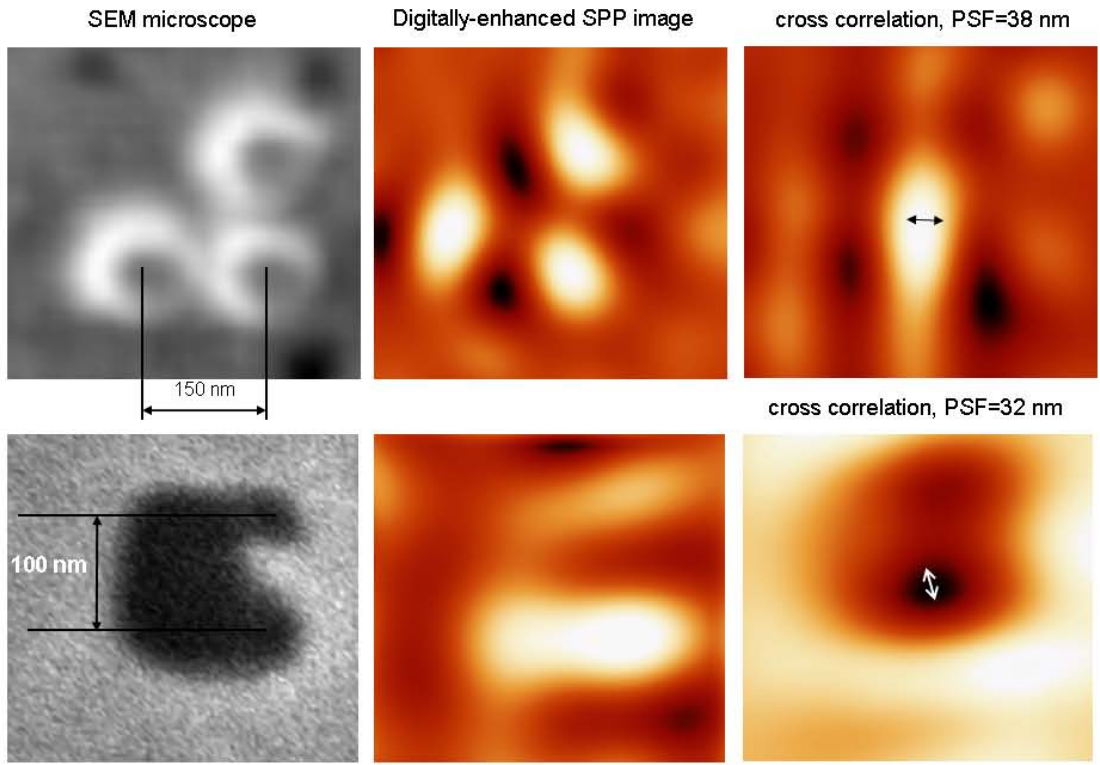

**Fig.6**